\documentclass[conference]{IEEEtran}
\usepackage{setspace}
\usepackage{amsmath}
\usepackage{acronym}
\usepackage{bbm}
\usepackage[dvips]{color}
\usepackage{comment}
\usepackage{todonotes}
\usepackage{epsf}
\usepackage{siunitx}
\usepackage{epsfig}
\usepackage{caption}
\usepackage{times}
\usepackage{epsfig}
\usepackage{graphicx}
\usepackage{bbold}
\usepackage{geometry}
\usepackage{mathrsfs}
\usepackage{amssymb}
\usepackage{pdfpages}
\usepackage{epstopdf}
\usepackage{tcolorbox,bbold}
\usepackage{algorithm,algorithmicx,algpseudocode}
\makeatletter
\newcommand{\algmargin}{\the\ALG@thistlm}
\makeatother
\algnewcommand{\parState}[1]{\State%
  \parbox[t]{\dimexpr\linewidth-\algmargin}{\strut #1\strut}}

\usepackage[nomain,acronym,shortcuts]{glossaries}
\newcommand*{\acro}[3][]{\newacronym[#1]{#2}{#2}{#3}}

\acro{D2D}{device-to-device}
\acro{SIR}{signal-to-interference-ratio}
\acro{SINR}{signal-to-interference-plus-noise-ratio}
\acro{PCP}{Poisson cluster process}
\acro{CoMP}{coordinated multi-point}
\acro{BS}{base station} 
\acro{MD-CoMP}{macrodiversity CoMP transmission}
\acro{MAC}{medium-access-control}
\acro{JT-CoMP}{joint transmission CoMP}
\acro{CoMP-JT}{coordinated multipoint joint transmission}
\acro{SBS}{small base station}
\acro{MDSD}{multiple devices to a single device}
\acro{MDS}{maximum distance separable}
\acro{SCN}{small cell network}
\acro{PPP}{Poisson point process}
\acro{TCP}{Thomas cluster process}
\acro{CSI}{channel state information}
\acro{PDF}{probability distribution function}
\acro{PMF}{probability mass function}
\acro{RV}{random variable}
\acro{i.i.d.}{independently and identically distributed}
\acro{MBMS}{multimedia broadcasting multicasting service}
\acro{EE}{energy efficiency}
\acro{HCP}{hard-core placement}
\acro{CCDF}{complementary cumulative distribution function}
\acro{CDF}{cumulative distribution function}
\acro{PC}{probabilistic caching}
\acro{RC}{random caching}
\acro{CPF}{caching popular files} 
\acro{PGFL}{probability generating functional}
\acro{KKT}{Karush-Kuhn-Tucker}
\acro{PGF}{point generating function}
\acro{SCA}{successive convex approximation}
\acro{HD}{high-definition}
\acro{FHD}{full-high-definition}
\acro{UHD}{ultra-high-definition}
\acro{VR}{virtual reality}
\acro{AR}{augmented reality}
\acro{5G}{fifth-generation}
\acro{QoS}{quality-of-service}
\acro{QoE}{quality-of-experience}
\acro{IoT}{internet of things}
\acro{MHCPP}{Matern hardcore point process}
\acro{LoS}{line-of-sight}
\acro{NLoS}{non-line-of-sight}
\acro{PSD}{power spectral density}
\acro{MEC}{mobile edge computing}
\acro{E2E}{end-to-end}
\acro{THz}{terahertz}
\acro{CLT}{central limit theorem}
\acro{HQ}{High Quality}
\acro{eMBB}{enhanced mobile broadband}
\acro{URLLC}{ultra reliable low latency communications}
\acro{mmWave}{millimeter wave}
\acro{EVT}{extreme value theory}
\acro{GEV}{generalized extreme value}
\acro{LIS}{large intelligent surface}
\acro{RIS}{reconfigurable intelligent surface}
\acro{RF}{radio frequency}
\acro{UE}{user equipment}
\acro{MIMO}{multiple-input multiple-output}
\acro{EVaR}{entropic value-at-risk}
\acro{DNN}{deep neural network}
\acro{MDP}{Markov decision process}
\acro{RL}{reinforcement learning}
\acro{RNN}{recurrent neural network}
\acro{ANN}{artificial neural networks}
\acro{LSTM}{long short-term memory}
\acro{ReLu}{rectified linear unit}
\acro{VaR}{value-at-risk}
\acro{SNR}{signal-to-noise ratio}

\usepackage{amssymb}
\usepackage{url}
\usepackage{dsfont}
\usepackage{lettrine} 
\usepackage{amsmath,epsfig,amssymb,algorithm,algpseudocode,amsthm,cite,url}
\IEEEoverridecommandlockouts
\usepackage{subcaption}
\allowdisplaybreaks
\usepackage{csquotes}
\usepackage[english]{babel}
\usepackage{amsmath,amssymb}

\usepackage[justification=centering]{caption}
\usepackage{verbatim}

\newtheorem{proposition}{\bf Proposition}

\IEEEoverridecommandlockouts
\begin{document}
	\title{Risk-Based Optimization of Virtual Reality over Terahertz Reconfigurable Intelligent Surfaces \vspace{-0.35cm}}
	\author{\IEEEauthorblockN{Christina Chaccour\IEEEauthorrefmark{1},
		Mehdi Naderi Soorki\IEEEauthorrefmark{2},
		Walid Saad \IEEEauthorrefmark{1},
		Mehdi Bennis\IEEEauthorrefmark{3}, 
	and Petar Popovski \IEEEauthorrefmark{4}}
	\IEEEauthorblockA{\IEEEauthorrefmark{1}Wireless@ VT, Bradly Department of Electrical and Computer Engineering, Virginia Tech, Blacksburg, VA USA,}
	\IEEEauthorblockA{\IEEEauthorrefmark{2}Faculty of Engineering, Shahid Chamran University of Ahvaz, Ahvaz, Iran,}
	\IEEEauthorblockA{\IEEEauthorrefmark{3}Centre for Wireless Communications, University of Oulu, Finland,}
	\IEEEauthorblockA{\IEEEauthorrefmark{4} Department of Electronic Systems, Aalborg University, Denmark.}
	\IEEEauthorblockA{Emails:\{christinac, mehdin, walids\}@vt.edu, mehdi.bennis@oulu.fi, petarp@es.aau.dk }
	\thanks{This research was supported by the U.S. National Science Foundation under Grant CNS-1836802, and in part, by the Academy of Finland Project CARMA, by the Academy of Finland Project MISSION, by the Academy of Finland Project SMARTER, as well as by the INFOTECH Project NOOR.}
	\vspace{-1cm}}
	\maketitle
\begin{abstract}
In this paper, the problem of associating \acp{RIS} to \ac{VR} users is studied for a wireless \ac{VR} network. In particular, this problem is considered within a cellular network that employs \ac{THz} operated \acp{RIS} acting as base stations. To provide a seamless \ac{VR}  experience, high data rates and reliable low latency need to be continuously guaranteed. To address these challenges, a novel risk-based framework based on the entropic value-at-risk is proposed for rate optimization and reliability performance. Furthermore, a Lyapunov optimization technique is used to reformulate the problem as a linear weighted function, while ensuring that higher order statistics of the queue length are maintained under a threshold. To address this problem, given the stochastic nature of the channel, a policy-based \ac{RL} algorithm is proposed. Since the state space is extremely large, the policy is learned through a deep-\ac{RL} algorithm. In particular, a \ac{RNN} \ac{RL} framework is proposed to capture the dynamic channel behavior and improve the speed of conventional \ac{RL} policy-search algorithms. Simulation results demonstrate that the maximal queue length resulting from the proposed approach is only within $1\%$ of the optimal solution. The results show a high accuracy and fast convergence for the \ac{RNN} with a validation accuracy of $91.92\%$.
\end{abstract}
\vspace{0cm}
\section{Introduction}\label{sec:Intro}
\Acf{VR} applications will revolutionize the way in which humans interact by allowing them to be immersed, in real time, in a range set of virtual environments \cite{minzghe}. Nevertheless, unleashing the potential of VR requires their integration into wireless networks in order to provide a seamless and immersive VR experience \cite{saad2019vision}.
However, deploying wireless VR services faces many technical challenges, the most fundamental of which is providing high rate wireless links with high reliability. On the one hand, \ac{VR} communication requires high data rates to guarantee a seamless visual experience while delivering $360^\circ$ \ac{VR} content.  On the other hand, providing reliable haptic VR communications will also require maintaining a very low \ac{E2E} delay in face of extreme and uncertain network conditions.\\
\indent Guaranteeing this dual performance requirement constitutes a major departure from classical \ac{URLLC} services limited to low-rate sensors \cite{urllconly}, or traditional \ac{eMBB} services  limited to high capacity delivered to dense networks \cite{steeg2017high}. In order to overcome the rate challenge of VR communications, one can explore the high bandwidth available at the \acf{THz} frequency bands \cite{coverage}. However, the reliability of the \ac{THz} channel can be impeded by its susceptibility to blockage, molecular absorption, and communication range. This, in turn, can violate the reliability requirements of VR systems. In order to alleviate these reliability concerns, one can deploy \acfp{RIS} \cite{basar2019wireless} acting as a \ac{BS} that can provide a nearly continuous line-of-sight (LoS) connectivity to VR users.  The \ac{RIS} concept can be viewed as a scaled-up version of conventional \ac{MIMO} systems beyond their traditional large array concept, however, an \ac{RIS} exhibits several key differences from massive \ac{MIMO} systems \cite{basar2019wireless}. Most fundamentally, \acp{RIS} will be densely located in both indoor and outdoor spaces, making it possible to perform near-field communications through a \ac{LoS} path. Hence, coupling \acp{RIS} with \ac{THz} communications can potentially provide connectivity that exhibits both high data rates and high reliability (in terms of guaranteeing \ac{LoS} communication). Moreover, \ac{VR} users will always be at a proximity of physical structures with high rate wireless capabilities. Thus, it is imperative to understand whether \ac{THz}-operated \acp{RIS} can indeed provide an immersive \ac{VR} experience by delivering continuously reliable connectivity with low \ac{E2E} delay and high data rate.\\
\indent A number of recent works attempted to address the challenges of \ac{VR} communications \cite{kasgari2019human, minzghe, chen2019deep, urllcembb}. In \cite{kasgari2019human}, the authors study the spectrum resource allocation problem with a brain-aware \ac{QoS} constraint. The work in \cite{minzghe} proposes a \ac{VR} model that captures the tracking and delay components of VR \ac{QoS}. Meanwhile, the work in \cite{chen2019deep} proposes a novel framework that uses
cellular-connected drone aerial vehicles to collect \ac{VR} content for reliable wireless transmission. In \cite{urllcembb} the authors study the issue of concurrent support of visual and haptic perceptions over wireless cellular networks. However, the works in \cite{minzghe} and \cite{kasgari2019human, urllcembb, chen2019deep} do not account for realistic delays and their statistics, and their solutions cannot satisfy high rates and low latency simultaneously. In contrast, to provide reliable \ac{VR}, it is of interest to explore the possibility of deploying \acp{RIS}. If properly operated, serving \ac{VR} users through existing walls and structures with wireless capabilities will unleash the potential of reliable \ac{VR}. Specifically, equipping \acp{RIS} with \ac{THz} will guarantee the overall seamless experience. We also note that despite the surge of recent works on \ac{THz} communications (e.g., see \cite{softwaredefined} and \cite{coverage}, and references therein) and \ac{RIS} design and optimization (e.g., see \cite{huang2019reconfigurable, jung2019uplink}, and references therein), these works focus on the physical layer and do not address \ac{VR} or networking challenges of \ac{THz} communications.\\
\indent The main contribution of this paper is a novel rate and reliability optimization framework for VR systems leveraging \ac{THz}-operated \acp{RIS}. We consider the downlink of a cellular network in which \ac{THz}-operated
\acp{RIS} serve \ac{VR} users. In this network, due to the mobility of users and the stochastic nature of the channel, the \acp{RIS} must be dynamically and intelligently scheduled to \ac{VR} users. Also, to guarantee reliability and capture a full knowledge of delay statistics, we propose a novel approach that exploits the economic concept of \ac{EVaR} which \emph{coherently} measures the risk associated to a random event \cite{ahmadi2012entropic}. Hence, the EVaR is employed so as to capture higher order statistics of the delay and, thus, allowing us to define a concrete a measure of the risk associated to delay unreliability. We then formulate a high reliability and sum rate maximization scheduling problem by combining both Lyapunov optimization and \acp{DNN}.  Using the Lyapunov optimization technique, the problem is
transformed into a linear weighted function, which ensures that
the maximum queue length and the maximum queue length variance among \ac{VR} users remains bounded. To solve the proposed Lyapunov optimization problem, we propose a \acf{RL} algorithm based on \acfp{RNN} that can find the user associations to \acp{RIS} while capturing the dynamic temporal behavior of the users in the channel. Simulation results show that the gap between the proposed approach and the optimal solution is minimal.\\
\indent The rest of this paper is organized as follows. The system
model is presented in Section II. The risk aware association for \ac{VR} users is proposed in Section III. The \ac{RL} approach is presented Section IV.
In Section V, we provide simulation results.
Finally, conclusions are drawn in Section VI.
\vspace{-0.2cm}
\section{System Model}\label{Sec:Sys-Model}
\begin{figure}[!t]
	\centering
	\includegraphics[width=0.4\textwidth]{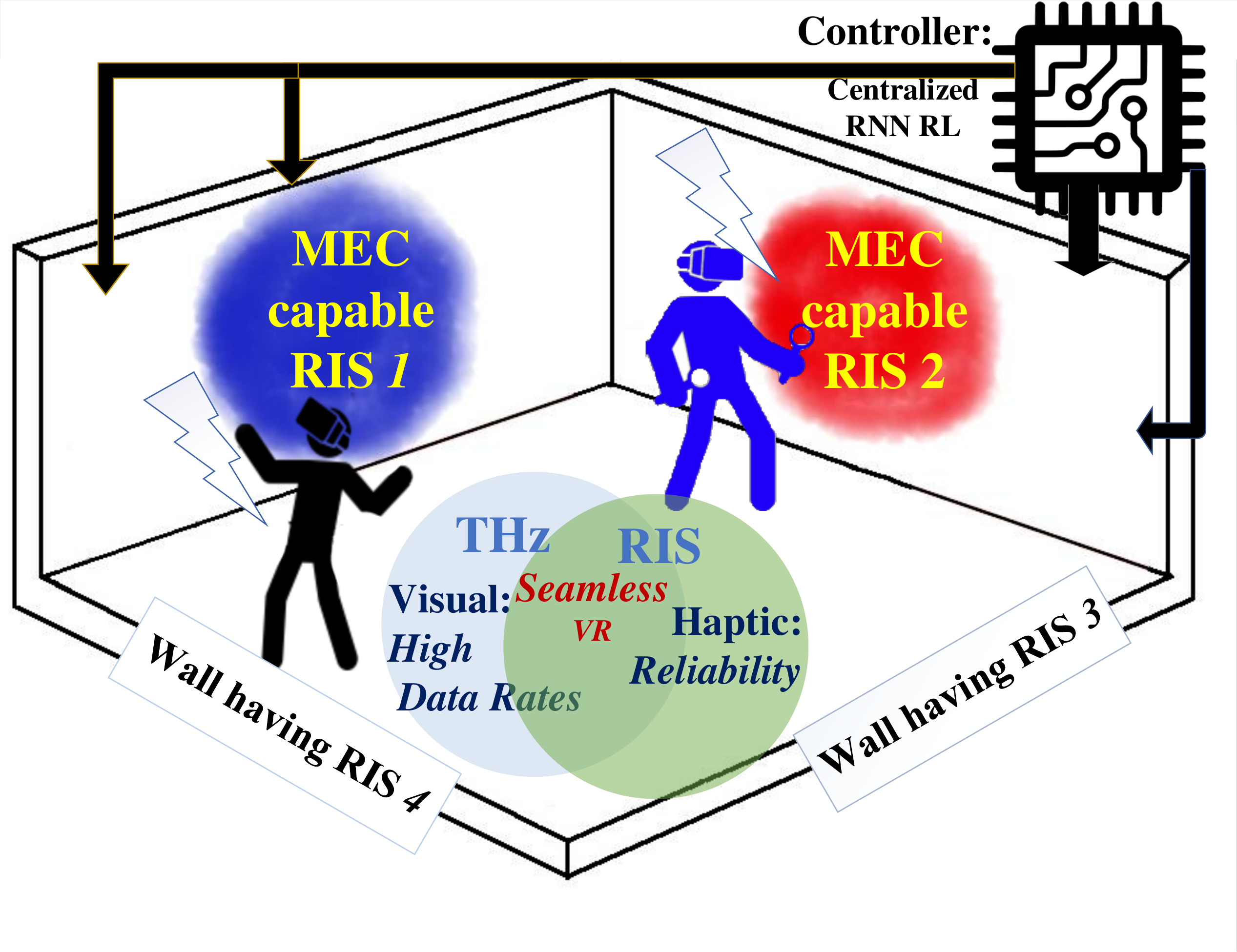}
	\caption{\small{Illustrative example of our system model.}}
	\label{fig:model}
	\vspace{-0.7cm}
\end{figure}

Consider the downlink\footnote{\noindent The uplink of VR requests is assumed to follow an arbitrary URLLC scheme and is outside of the scope of this paper} of an \ac{RIS}-based wireless network in a confined indoor area, servicing a set $\mathcal{U}$ of $U$ mobile wireless \ac{VR} users via a  set $\mathcal{B}$ of $B$ \acp{RIS} acting as \ac{THz} operated \acp{BS} as depicted in Fig.~\ref{fig:model}. 
The \ac{VR} users are mobile and may change their locations and orientations at any point in time. We consider discrete time slots indexed by $t$ with fixed duration $\tau$. Each \ac{RIS} is a \ac{BS}, that is provided with a feeder (antenna) with a corresponding transmit power denoted by $p$. Hence,  the transmitted data is encoded onto the phases of the signals reflected from different reconfigurable meta-surfaces that compose the \ac{RIS} \cite{basar2019wireless}. Henceforth, if the \ac{RIS} consists of $N$ meta-surfaces whose reflection phase can be optimized independently, then an $N$-stream virtual \ac{MIMO} system can be realized by using a single \ac{RF} active chain \cite{basar2019wireless}. We assume that the \ac{RF} source is close enough to the \ac{RIS} surface so that the transmission between each pair of \ac{RF} source and \ac{RIS} is not affected by fading. Then, the electromagnetic response of the $N$ meta-surface elements
can be programmed by using a centralized controller, which generates
input signals that tune varactors and change the phase of
the reflected signal~\cite{huang2019reconfigurable}. Let $\Phi_{bu,t}=[\phi_{bun,t}]_{N \times 1}$ be the phase shift vector of \ac{RIS} $b$ , with respect to the \ac{UE} $u$, at time slot $t$, where $\phi_{bun,t} \in \Phi$, $n$ is the index corresponding to the meta-surface of each \ac{RIS}, and $\Phi=\{-\pi+\frac{2z\pi}{Z-1}|z=0,1,...,Z-1\}$. $Z$ is the number of possible phase shifts per meta-surface element.
\vspace{-0.3cm}
\subsection{Channel Model}
\vspace{-0.1cm}
Due to the mobility of the \ac{VR} users, the \ac{THz} link between a \ac{VR} user and its respective \acp{RIS} may be blocked by self-blockage, i.e., the event of blocking the signal received by \ac{UE} $u$'s own body, or by dynamic blockages, i.e., blocking the signal by other \ac{VR} users' bodies respectively. Let $ {s}_{bu,t}$ be a random binary variable where $  s_{bun,t}=1$ if there is a \ac{LoS} link between \ac{RIS} $b$ and  \ac{VR} \ac{UE} $u$ at time slot $t$, and $  s_{bun,t}=0$, otherwise. As a byproduct of directional beamforming and propagation differences, the network considered is noise limited. Thus, the random channel gain between \ac{RIS} $b$ and \ac{VR} UE $u$ at time slot $t$ is given by \cite{chaccour2019reliability}:
\begin{equation*}
  {h}_{bu,t}=
  \begin{cases}
      \big(\frac{\lambda}{4\pi d_{bu,t}}\big)^2\big(e^{-k(f)   d_{bu,t}}\big)^2,& \quad \text{with } \Pr(  {s}_{bun,t}=1),\\
		0,  & \quad \text{with } \Pr(  {s}_{bun,t}=0).\\
  \end{cases}
  \label{Channel_gain}
\end{equation*}
$d_{bu,t}$ is the distance between \ac{RIS} $b$ and the \ac{VR} \ac{UE} $u$ at time slot $t$, $k(f)$ is the overall molecular absorption coefficients of the medium at \ac{THz} band, and $f$ is the operating frequency.
\noindent Let $\psi_{bun,t}$ be the phase shift of the channel between  \ac{VR} \ac{UE} $u$ and the meta-surface $n$ of \ac{RIS} $b$ at $t$. Then, for a given reflection phase shift vector, $\Phi_{bu,t}$, the transmission rate from \ac{RIS} $b$ to \ac{VR} UE $u$ will be (under an approximate average \ac{SNR} value across the \ac{THz} band):
\vspace{-0.1cm}
\begin{equation}
  {\tilde r}_{bu,t}= W \log_2 \left( 1+\frac{p   {h}_{bu,t} |\sum_{n=1}^{N} e^{(\phi_{bun,t}-\psi_{bun,t})j}|^2  {s}_{bun,t}}{N({  d_{bu,t}},p,f)} \right) ,
  \label{Received bitrate}
\end{equation}
where $N({  d_{u,t}},p,f)=N_0+\sum_{b=1}^{B}pA_0d_{bu,t}^{-2}(1-e^{-K(f)  d_{bu,t}})$,  $N_0=\frac{W \lambda^2}{4 \pi} k_B T_0$, $k_B$ is the Boltzmann constant, $T_0$ is the temperature in Kelvin,  $A_0=\frac{c^2}{16 {\pi}^2 f^2}$, and $c$ is the speed of light \cite{zhang2018analytical, chaccour2019reliability}. Note that the optimal choice for $\phi_{bun,t}$ for every \ac{RIS} association is equal to $\psi_{bun,t}$, thus maximizing the rate ${\tilde r}_{bu,t}$, as shown in \cite{basar2019wireless}. This selection will be made by the controller after learning the \ac{RIS} association and optimizing it, as shown in subsequent sections.
\subsection{Queuing Model}
Each \ac{RIS} is equipped with \ac{MEC} capabilities, and, thus, we model the queuing and transmission of each \ac{VR} content as an M/G/1 queue at the \ac{MEC} server of the \ac{RIS}. We define a decision binary variable $x_{bu,t}$ that is equal to 1 if \ac{RIS} $b$ is scheduled to serve the \ac{VR} content queue of user $u$ at time slot $t$, otherwise $x_{bu,t}=0$. Note that, multiple users can be associated to one \ac{RIS}, however, each user is connected to a single RIS.
Let $  Q_u(t)$ be the queue length corresponding to \ac{UE} $u$'s requested \ac{VR} image at the beginning of slot $t$, then the queue dynamics are given by:
\begin{equation}
  {Q}_{u,(t+1)}=\max\{   Q_{u,t}-  {\tilde R}_{bu,t},0\}+  {A}_{u,t}, \text{if } x_{bu,t}=1,
\label{queue}
\end{equation}
where $A_{u,t}$ is the number of \ac{VR} images queued for transmission at time slot $t$. The arrival of \ac{VR} content follows a Poisson arrival process with mean rate $\lambda_{u}$. $  {\tilde R}_{bu,t}$ is the rate of \ac{VR} image transmission over \ac{THz} link between \ac{RIS} $b$ and \ac{VR} \ac{UE} $u$ at time slot $t$. $ {\tilde R}_{bu,t}=\frac{  {\tilde r}_{bu,t}\tau}{M}$ where $M$ is the size of the \ac{VR} image. Given that the availability of the \ac{THz} \ac{LoS} link is a random variable, $  {\tilde R}_{bu,t}$ is a stochastic random variable with respect to time. 
\section{Risk-Aware RIS-VR user Association}
\subsection{Problem Formulation}
Our goal is to characterize the \ac{RIS}-\ac{UE} association policy which determines the system parameters over a finite horizon of length $T$. The objective of this optimal policy is to maximize the sum-rate while maintaining  reliable transmission. Formally, we define a \emph{policy} $\Pi_{t}=\{x_{bu,t}|\forall b \in \mathcal{B},\forall u \in \mathcal{U}\}$ for the controller that associates each \ac{RIS} to its respective \ac{VR} users. The control policy at a given slot $t$ depends on unknown environmental changes, which is a consequence of the stochastic nature of the channel and the sudden changes that might block the \ac{LoS} signal between \ac{RIS}s and mobile \ac{VR} \ac{UE}. Thus, 
$\Pr(  s_{bun,t}=j|  s_{bun,(t-1)}=i),\forall b \in \mathcal{B},\forall u \in \mathcal{U}$, over the \ac{LoS} \ac{THz} links.
Furthermore, the  reliability metric is satisfied as long as the \ac{CDF} of the \ac{E2E} delay does not exceed the reliability constraint associated with its respective network. Subsequently, to account for the risk of loss incurred when reliability is not satisfied,  the \ac{VaR} concept defined as VaR$_{1-\alpha}=- \inf_{t\in \mathbb{R}} \{P(X \leq t) \geq 1-\alpha\}$ \cite{ahmadi2012entropic}, can be used. However, \ac{VaR} is an uncoherent risk measure, making its analysis intractable. Thus, we define the \ac{EVaR} as $\phi_{t}= \frac{\log \mathbb {E} \left[ \exp(-\gamma   Q_{t})\right]}{\gamma}$, which is a coherent risk measure that corresponds to the tightest possible upper bound obtained from the \ac{VaR}. 
In the \ac{EVaR}, $   Q_t:=\max_{u \in \mathcal{U}} \{    Q_{u,t} \} $ and $0<\gamma\ll 1$. Subsequently, to ensure reliability, the following condition needs to be met: $\lim_{t\rightarrow\infty} \phi_{t} < \kappa$. Expanding the Maclaurin series of $\phi_t$ with respect to the $\log$ and $\exp$ functions we obtain, $\phi_t=E[   Q_t]+ \vartheta( Q_t)-1 +\mathcal{O}(   Q_t^3)$, where $\vartheta(Q_t)= E[   Q_t^2]- E[Q_t]^2$ is the variance of the maximum queue length. Thus, to minimize $\lim_{t\rightarrow\infty} \phi_{t}$, it is sufficient to minimize the first two terms of its Maclaurin series. Consequently, we formulate the \ac{RIS} association and phase shift-control problem for an \ac{RIS}-assisted \ac{THz} indoor network as follows:
\begin{align}
&\underset{\left\{\substack{\Pi_t\substack}\right\}}
\max \sum_{b\in \mathcal{B}} \sum_{u\in \mathcal{U}} x_{bu,t}   {\tilde R}_{bu,t} ,\label{opt_problem1}\\
\text{s.t.} &  \lim_{T \rightarrow\infty} \frac{1}{T} \sum_{t=1}^{T} \mathbb{E}[  {Q}_{t}] < \varepsilon, \label{opt_problem1_const1}\\
&  \lim_{T \rightarrow\infty} \frac{1}{T} \sum_{t=1}^{T} \mathbb{E}[  {Q}_{t}^2] < \eta, \label{opt_problem1_const2}\\
& \phi_{bu,t} \in \Phi, \forall{b}\in\mathcal{B},\forall{u}\in\mathcal{U}, \forall{t}\in\mathcal{T},\label{opt_problem1_const3}\\
& \sum_{u \in \mathcal{U}}x_{bu,t}\leq 1, \forall{b}\in\mathcal{B}, \forall{t}\in\mathcal{T}, \label{opt_problem1_const4}\\
& x_{bu,t}\in \{0,1\}, \forall{b}\in\mathcal{B},\forall{u}\in\mathcal{U}, \forall{t}\in\mathcal{T}. \label{opt_problem1_const5}
\end{align}
Here, maximizing the objective function in \eqref{opt_problem1} ensures that the visual component of the \ac{VR} experience is guaranteed, thus, delivering a seamless experience. On the other hand, \eqref{opt_problem1_const1} and \eqref{opt_problem1_const2} ensure that the constraint of mitigating the risk will be satisfied, where $\eta=\varepsilon^2 +2 \left[ \gamma(\kappa+1) - \varepsilon\right]$, this further guarantees that the haptic component of the \ac{VR} will be delivered successfully. Given that the length of the queues changes with random events and their probability distribution is not known a priori, the optimization problem in \eqref{opt_problem1} cannot be solved using traditional stochastic optimization techniques \cite{neely2010stochastic}. Next, we propose a tunable minimum-drift-plus-penalty optimization problem based on Lyapunov optimization to reformulate the problem stated previously.
\vspace{-0.25cm}
\subsection{Lyapunov Optimization}
We use a Lyapunov optimization approach \cite{neely2010stochastic} to solve
problem \eqref{opt_problem1}. This approach allows us to convert the constraints into a tractable form. Henceforth, to ensure \eqref{opt_problem1_const1} and \eqref{opt_problem1_const2}, we define two virtual queues $Z_1$ and $Z_2$, having with the following dynamics:
\vspace{-0.35cm}
\begin{align}
	   Z_{1,(t+1)}=\max\{ Z_{1,t} +    Q_t -\varepsilon,0\},\\
	   Z_{2,(t+1)}=\max\{ Z_{2,t}+    Q^2_{t} -\eta,0\}.
\end{align}
Moreover, given that our initial optimization problem is a maximization problem, our aim is to minimize the drift-plus-penalty expression given by:	
\begin{equation}
\Delta_t -V\sum_{b\in \mathcal{B}} \sum_{u\in \mathcal{U}} x_{bu,t}   {\tilde R}_{bu,t},
\end{equation}
where 	$\Delta_t=\mathbb{E}[ L_{t+1}- L_{t}|    Q_t]$, $  {L}_{t}$ is the Lyapunov function given by $  L_t=\frac{1}{2}(Z^2_{1,t}+Z^2_{2,t}+\sum_{u \in \mathcal{U}}   {Q}^2_{u,t})$. Next, we transform problem \eqref{opt_problem1} into one whose objective is a linear weighted function, and its 
constraints are no longer a function of $Q_t$ as in \eqref{opt_problem1_const1} and \eqref{opt_problem1_const2}.
\vspace{0.25cm}
\begin{proposition}
The conditional Lyapunov drift-plus-penalty bound under any feasible control policy $\pi_t$ is formulated as follows:
\vspace{-0.5cm}
\begin{align}
\label{eq::Lemmaaa}
	\Delta_t \leq & \Upsilon + \sum_{u=1} ^ U  {Q}_{u,t}(  {A}_{u,t}-  R_{bu,t})+   Z_{1,t}(   Q_t- \varepsilon) \nonumber \\
	& +   Z_{2,t}(   Q^2_t- \eta) -V\sum_{b\in \mathcal{B}} \sum_{u\in \mathcal{U}} x_{bu,t}   {\tilde R}_{bu,t}
\end{align}
\end{proposition}
	\vspace{-0.25cm}
\begin{IEEEproof}
	Given that for $\forall	 x \in \mathbb{R}, \max\{x,0\}^2 \leq x^2$, we subtract $Q_{u,t}$ on both sides and square \eqref{queue} as follows:
	\begin{align*}
	 Q^2_{u,t+1}-Q^2_{u,t} \leq& (Q_{u,t}- \tilde R_{bu,t})^2 +    A_{u,t}^2 +2    A_{u,t}(   Q_{u,t}-\tilde R_{bu,t})  \nonumber \\
	&-  {Q}^2_{u,t}.
	\end{align*}
	Simplifying the equation leads to the following:
	\vspace{-0.25cm}
	\begin{align*}
	\frac{Q^2_{u,t+1}-   Q^2_{u,t}}{2} \leq&  \frac{\left(\tilde 	 R_{bu,t}-   A_{u,t}\right)^2 }{2} +  Q_{u,t}(A_{u,t}-\tilde R_{bu,t}) 
	\end{align*}
	Similarly,
	\begin{align*}
	\frac{Z^2_{1,{t+1}}-Z^2_{1,t}}{2} \leq& 	\frac{\left( Q_t-\varepsilon \right)^2 }{2} +   Z_{1,t}(Q_t-\varepsilon),\\
	\frac{Z^2_{2,{t+1}}-   Z^2_{2, t}}{2} \leq& 	\frac{\left( 	   Q^2_{t}-\eta \right)^2 }{2} +   Z_{2,t}(Q^2_{t}-\eta).
	\end{align*}
	After some mathematical manipulation, we obtain:
	\begin{align}
	L_{t+1}-L_{t} \leq & \Upsilon + \sum_{u=1} ^ U  {Q}_{u,t}(  {A}_{u,t}-  {\tilde R_{bu,t}})+   Z_{1,t}(Q_t- \varepsilon) \nonumber \\
	&+Z_{2,t}(Q^2_t- \eta).
	\end{align}
	where $\Upsilon= \frac{U \left( \max_{b \in \mathcal{B},u \in \mathcal{U},t}  \tilde R_{bu,t}\right)^2 + \varepsilon^2 +\eta^2}{2}.$
\end{IEEEproof}
Thus, instead of minimizing the drift-plus-penalty expression, we minimize the maximum bound of the one-time slot conditional Lyapunov drift plus penalty in \eqref{eq::Lemmaaa}. The initial optimization problem is reformulated as:
\begin{align}
&\underset{\left\{\substack{\Pi_t\substack}\right\}}
\max  V\sum_{b\in \mathcal{B}} \sum_{u\in \mathcal{U}} x_{bu,t}   {\tilde R}_{bu,t}-
\sum_{u=1} ^ U  {Q}_{u,t}(  {A}_{u,t}-  \tilde R_{bu,t})- \nonumber \\
&   Z_{1,t}(   Q_t- \varepsilon)-    Z_{2,t}(   Q^2_t- \eta)
,\label{opt2_problem}\\
& \text{subject to}\nonumber \\
& \phi_{bu,t} \in \Phi, \forall{b}\in\mathcal{B},\forall{u}\in\mathcal{U}, \forall{t}\in\mathcal{T},\label{opt2_problem1_const3}\\
& \sum_{u \in \mathcal{U}}x_{bu,t}\leq 1, \forall{b}\in\mathcal{B}, \forall{t}\in\mathcal{T}, \label{opt2_problem1_const4}\\
& x_{bu,t}\in \{0,1\}, \forall{b}\in\mathcal{B},\forall{u}\in\mathcal{U}, \forall{t}\in\mathcal{T}. \label{opt2_problem1_const5}
\end{align}
Thus, employing virtual queues and Lyapunov optimization allowed us to transform our optimization problem into a linear weighted function. Nevertheless, solving problem \eqref{opt2_problem} using integer programming will be very complex due to its combinatorial nature and the stochasticity of the variables. Given that the distribution of system parameters is not characterizable, the problem cannot be solved using stochastic matching theory or stochastic optimization. Next, to solve \eqref{opt2_problem} we propose a centralized and low-complexity \ac{RNN} \ac{RL} framework that provides the optimal policy for the \ac{RIS}-\ac{UE} association. Moreover, \ac{RNN} \ac{RL} is suitable for this problem since it can reduce the dimensionality of the large state space, while capturing dynamic temporal behaviors \cite{chen2019artificial}.
\vspace{-0.25cm}
\section{Recurrent Neural Networks RL for Wireless VR in THz operated RIS network}
In this section, an adaptive control policy based on a deep \ac{RL} framework is proposed. The proposed framework will allow us to learn the policy to solve the problem of \ac{RIS}-\ac{UE} associations in \eqref{opt2_problem}. We model (\ref{opt2_problem}) as a \ac{MDP} represented by the tuple $\{\mathcal{S},\mathcal{A},P,R\}$, where $\mathcal{S}$ is the state space, $\mathcal{A}$ is the action space, $P$ is an unknown state transition function, $P(\boldsymbol{s}',\boldsymbol{s},\boldsymbol{a}) = \Pr(\boldsymbol{s}_{t+1}= \boldsymbol{s}'\mid \boldsymbol{s}_t = \boldsymbol{s}, \boldsymbol{a}_t = \boldsymbol{a})$, and $R(\boldsymbol{a}_t,\boldsymbol{s}_t)$ is the reward function~\cite{sutton2018reinforcement}. Our action space is the set of all possible \ac{RIS} associations  \ac{VR} \acp{UE}, $\mathcal{A}=\{ [x_{bu}]_{B\times u} \mid x_{bu}\in \{0,1\}, b=1,...,B, u=1,...,U\}$ and the  reward is  $R(\boldsymbol{a}_t,\boldsymbol{s}_t)=V\sum_{b\in\mathcal{B}} \sum_{u\in \mathcal{U}} x_{bu,t}   {\tilde R}_{bu,t}-\sum_{u=1} ^ U  {Q}_{u,t}(  {A}_{u,t}-  R_{bu,t})-  Z_{1,t}(   Q_t- \varepsilon)-   Z_{2,t}(   Q^2_t- \eta)$ which is the current objective function in  (\ref{opt2_problem}). The state is the set of \ac{VR} \ac{UE} queue lengths, virtual queue lengths,  and the state of \ac{LoS} links between \acp{RIS} and \ac{VR} \acp{UE}, $\mathcal{S}=\{ [s_{bu}]_{B\times u}, [Q_u]_{U\times 1},  Z_1, Z_2, \mid s_{bu}\in \{0,1\}, \{Q_{u},Z_1, Z_2\} \in \{\mathbb{Z}^+, b=1,...,B, u=1,...,U\}$.\\
\indent We represent the class of parameterized policies of our \ac{MDP} as $\Pi_t=\{ \pi_{\boldsymbol{\theta}}(\boldsymbol{a}_t|\boldsymbol{s}_t) \mid \theta \in \mathbb{R}^m \}$, where $\pi_{\boldsymbol{\theta}}(\boldsymbol{a}_t|\boldsymbol{s}_t)=\Pr\{\boldsymbol{a}=\boldsymbol{a}_t| \boldsymbol{s}=\boldsymbol{s}_t,\boldsymbol{\theta}\}$.
The stochastic reward function $R(\boldsymbol{a}_t,\boldsymbol{s}_t)$ during next time slot has a transition probability of  $\boldsymbol\rho_t=\prod_{\boldsymbol{s}'\in \mathcal{S}}\pi_{\boldsymbol{\theta}}(\boldsymbol{a}_t|\boldsymbol{s}_t)\Pr(\boldsymbol{s}_{t+1}= \boldsymbol{s}'|\boldsymbol{s}_t = \boldsymbol{s}, \boldsymbol{a}_t = \boldsymbol{a})$.
To solve the optimization problem in (\ref{opt2_problem}), the controller needs to have full knowledge about the transition probability and all possible values of $R(\boldsymbol{a}_t,\boldsymbol{s}_t)$ for all possible states of \ac{MDP} under a given policy $\pi_{\boldsymbol{\theta}}$.
Given that our model is highly dynamic due the mobility of users and the nature of the channel, the transition probability of states cannot be characterized through \acp{PDF}. Thus, it is necessary to use an \ac{RL} framework to solve (\ref{opt2_problem}).
We particularly use a \emph{policy-search} approach to find the optimal \ac{RIS} to \ac{VR} user association while maintaining a high reliability and a high data rate formulated in (\ref{opt2_problem}).
For each policy, we define its value as:
 \vspace{-0.5cm}
\begin{align}
 \label{eq::policy}
 J(\boldsymbol{\theta})=\sum&_{\boldsymbol{s}'\in \mathcal{S}} R(\boldsymbol{a}_t,\boldsymbol{s}_t) \boldsymbol{\rho_t}. 
\end{align}
Hence, to find the optimal policy, we need to find ${\theta}^*=\arg \max\limits_{\theta}J(\boldsymbol{\theta})$. To do so, we need to perform a gradient ascent on the policy parameters. Subsequently, we need to derive  $\nabla_{\boldsymbol{\theta}} J(\boldsymbol{\theta})$. Similarly to \cite{sutton2018reinforcement}, by writing  $\nabla_{\boldsymbol{\theta}}\log \pi_{\boldsymbol{\theta}}(\boldsymbol{a}_t|\boldsymbol{s}_t)=\frac{\nabla_{\boldsymbol{\theta}} \pi_{\boldsymbol{\theta}}(\boldsymbol{a}_t|\boldsymbol{s}_t)}{\pi_{\boldsymbol{\theta}}(\boldsymbol{a}_t|\boldsymbol{s}_t)}$ we can reformulate \eqref{eq::policy}, as follows:
\begin{equation}
\label{gradient}
	\nabla_{\boldsymbol{\theta}}J(\boldsymbol{\theta})\approx
	\mathbb{E}_{\Lambda_{t}}\{\nabla_{\boldsymbol{\theta}}\log \pi_{\boldsymbol{\theta}}(\boldsymbol{a}_t|\boldsymbol{s}_t) R(\boldsymbol{a}_t,\boldsymbol{s}_t)  \},
\end{equation}
where  $\Lambda_{t}=\{\boldsymbol{a}_{t}, \boldsymbol{s}_{t+1}= \boldsymbol{s}' \mid \boldsymbol{s'} \in \mathcal{S} \}$ is the trajectory of the \ac{MDP} for the next time slot. 
Subsequently, we can use \eqref{gradient} to solve the optimization problem in \eqref{opt2_problem} using a gradient ascent algorithm such as REINFORCE \cite{sutton2018reinforcement}. Nevertheless, given that the number of states is considerably high, this procedure will be intractable, which motivates the need for a function approximator through the use of \acp{DNN}. Given that the reliability depends on the prediction of the \ac{VR} users mobility pattern that will determine their associations to \acp{RIS}, it is important to implement a framework that is capable of capturing the dynamic behavior exhibited. To address these challenges, using an \ac{RNN} to represent the policy of \ac{RL} will extract the channel's dynamic features and learn an optimized sequence guaranteeing reliability at each time instant, based on the input features \cite{chen2019artificial}.
\begin{figure}[t]
		\centering
	\includegraphics[width=0.3\textwidth]{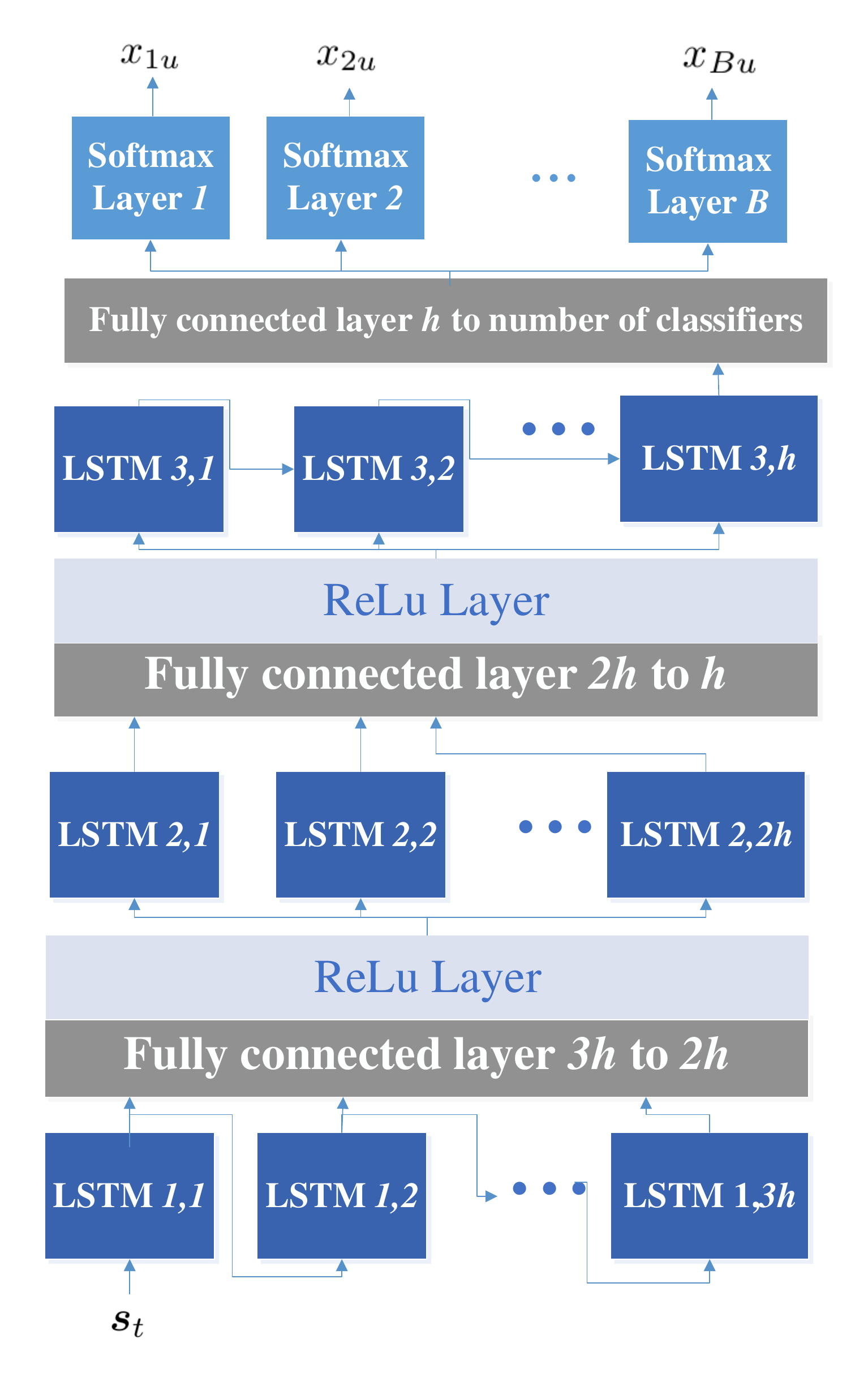}
	\caption{\small{Illustrative example of the RNN architecture.}}
	\label{fig:RNN}	
	\vspace{-0.5cm}
\end{figure} 
Since we deal with time-varying policies, it is natural to resort to \acp{RNN}. Indeed, \acp{RNN} are known to be effective in processing time-related data and capture dynamic temporal behaviors. As such, we represent the policy $\pi_{\boldsymbol{\theta}}(\boldsymbol a|\boldsymbol s)$ by an \ac{RNN}\cite{Naderi_2019} that can learn the user associations to \acp{RIS}. In particular, we use a many-to-many \ac{RNN} as shown in Fig.~\ref{fig:RNN}. The overall \ac{RNN} consists of an encoder and a decoder network: The encoder network comprises three \ac{LSTM} layers, two fully connected layers and two \ac{ReLu} layers. The state of the \ac{MDP} is thus encoded in the output of \ac{LSTM}  $3,h$. As for the decoder network, it consists of the last fully connected layer and the $B$ softmax layers. Moreover, the input consists of the states $\boldsymbol{s}_t \in \mathcal{S}$ of the \ac{MDP} that are fed to the first \ac{LSTM} layer of $h$ hidden layers. Subsequently, the $B$ softmax layers output the actions of the \ac{MDP} $a_t \in \mathcal{A}$, i.e., the $b$th softmax layer outputs $\{x_{bu} \mid x_{bu} \in {0,1}, b=1,\dots, B, u=1,\dots, U\}$. This architecture was chosen given that the \ac{LSTM} layers allow us to avoid the problem of vanishing gradients \cite{sutton2018reinforcement}; more precisely compared to other \acp{DNN}, it provides a faster \ac{RL} algorithm via a slow \ac{RL}. That is, instead of depending on the convergence speed of the \ac{RL} algorithm through conventional gradient ascent, its policy is represented by an \ac{RNN}. Subsequently, given that the \ac{RNN} receives all the typical information that a regular \ac{RL} algorithm would receive, the activations of the \ac{RNN} store the state that would improve the speed of the \ac{RL} algorithm on the current \ac{MDP} \cite{duan2016rl}.
\vspace{-0.2cm}
\section{Simulation Results and Analysis}
For our simulations, we consider the following parameters: $T= \SI{300}{K}$, $p=\SI{1}{W}$, $M= \SI{10}{Mbits}$, $f=\SI{1}{THz}$,  $W=\SI{30}{GHz}$, $K(f)= \SI{0.0016}{m^{-1}}$ with $1\%$ of water vapor molecules as in \cite{absorption}. The \acp{RIS} are deployed over the 4 walls of  an indoor area modeled as a square of size $\SI{40}{m}\times\SI{40}{m}$.  All statistical results are averaged over a large number of independent runs. In order to train the network, we consider the \ac{RNN} architecture shown in Fig. \ref{fig:RNN} with a maximum epoch of  $1000$ and a minimum batch size of $128$. Furthermore, the network was trained with data generated from \ac{VR} users moving according to a \emph{random walk} which constitutes the most general scheme characterizing users' mobility\footnote{Our approach can accommodate any other mobility model.} \cite{Naderi_2019}.
\begin{figure}[!t]
	\begin{center}
		\includegraphics[width=0.5\textwidth]{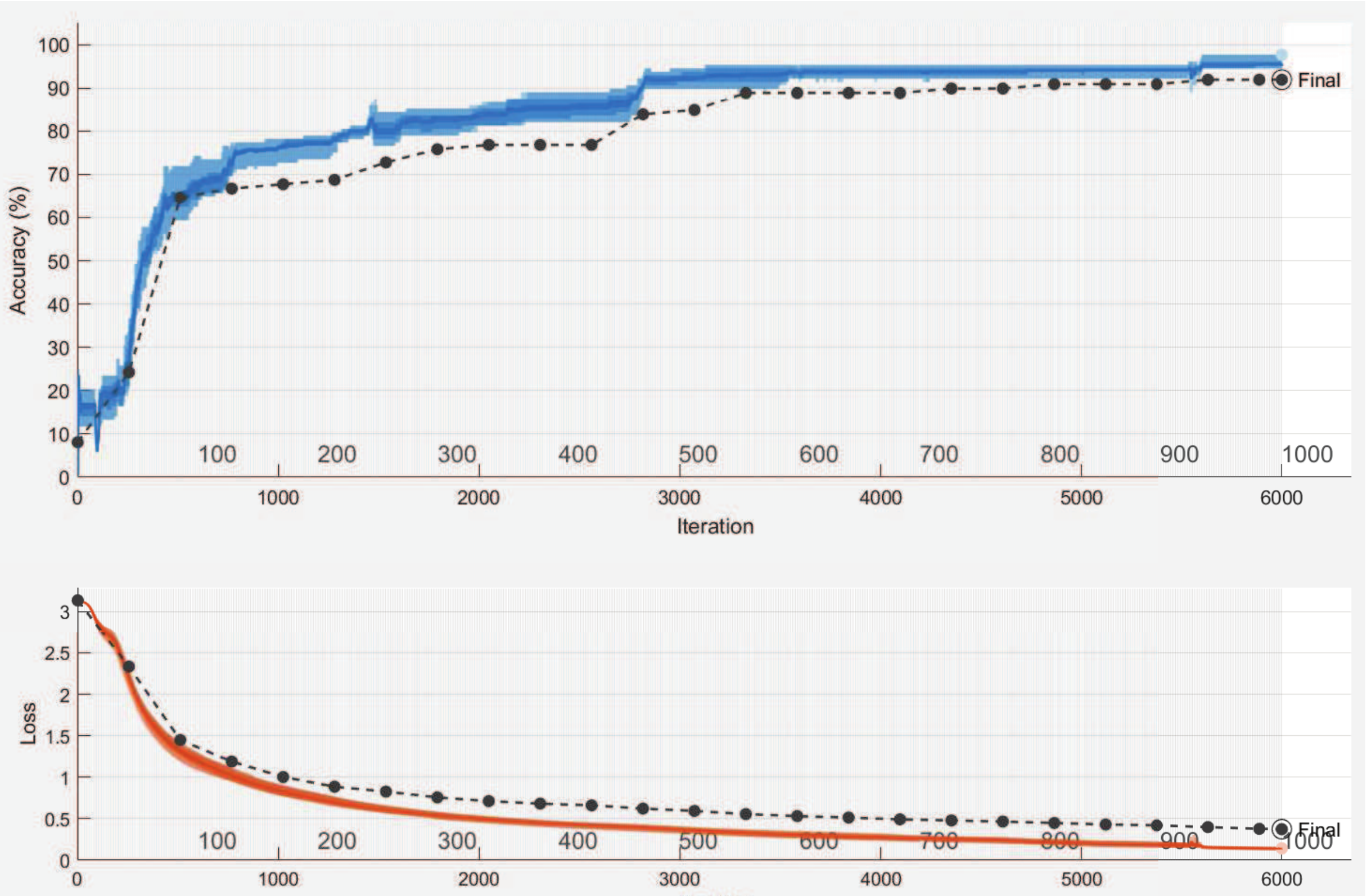}	\caption{ \small The training and validation process of the RNN.}
		\label{Training_Proc}
	\end{center}
\end{figure}
In Fig. \ref{Training_Proc}, we analyze the convergence of our proposed \ac{RNN}-\ac{RL} algorithm, in terms of accuracy and training loss. Note that, 80\% of our dataset is used for the training process,  10\% is used for the validation, and the remaining 10\% is used for testing. In particular, the training process uses data generated from users' random walks and the optimal solution to fit the model. Subsequently, throughout the validation process, the random walks' data is used to provide unbiased estimators of the hyper-parameters corresponding to the \ac{RNN}. Fig.~\ref{Training_Proc} shows a validation accuracy of $91.92\%$. Both the accuracy and loss of the training and validation processes show smoothness in the curve. Finally, after obtaining all the hyper-parameters of the \ac{RNN}, the test dataset can provide an unbiased evaluation of a final model fit. As such, simulation results show a testing error of $0.97\%$.\\
\begin{figure}[!t]
	\begin{center}
		\includegraphics[width=0.52\textwidth]{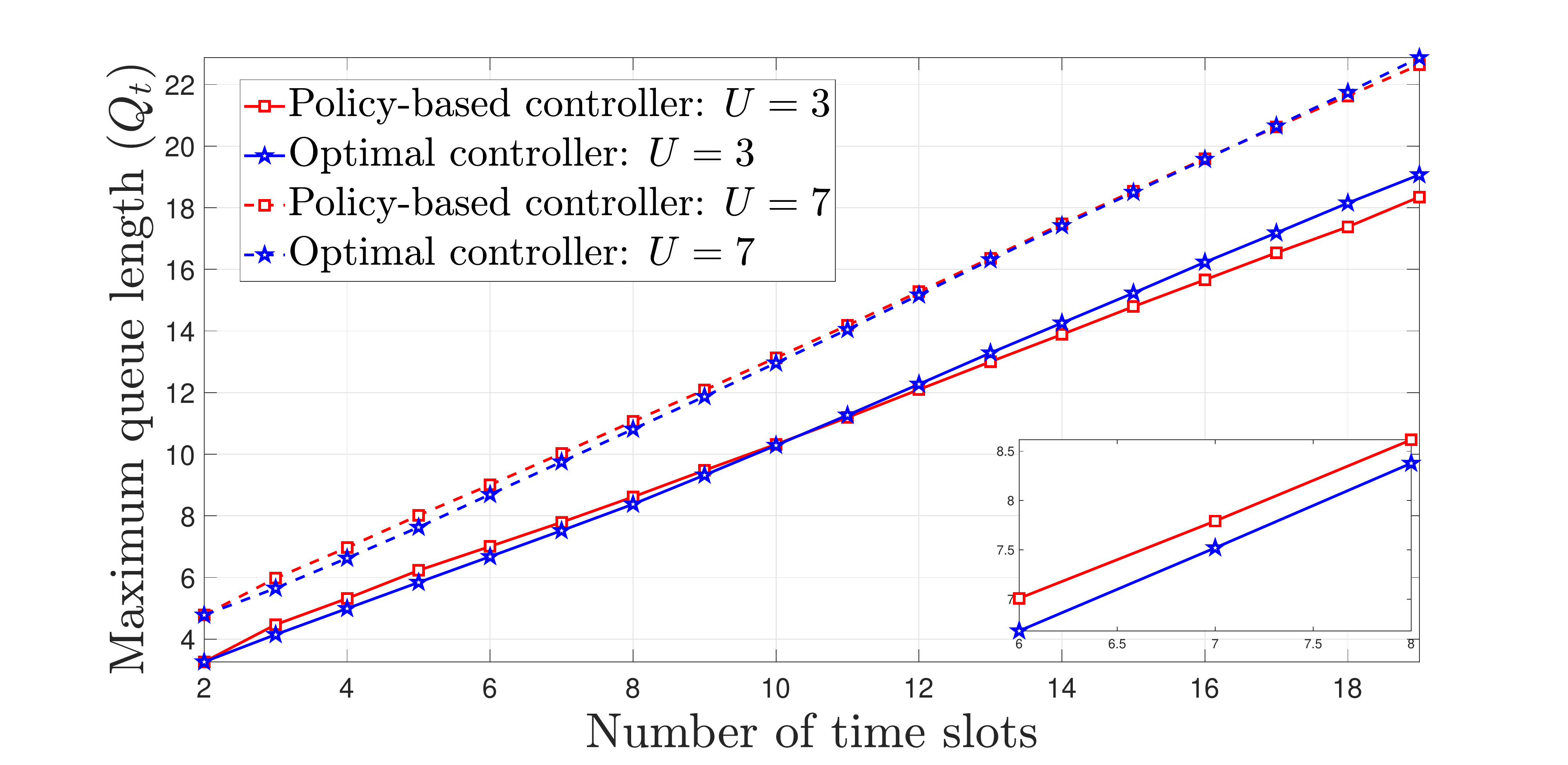}	
		\caption{ \small Maximum queue length vs. number of time slots.}
		\label{Maximum_Q}
	\end{center}
\vspace{-0.5cm}
\end{figure}
\begin{figure}[!t]
	\begin{center}
		\includegraphics[width=0.52\textwidth]{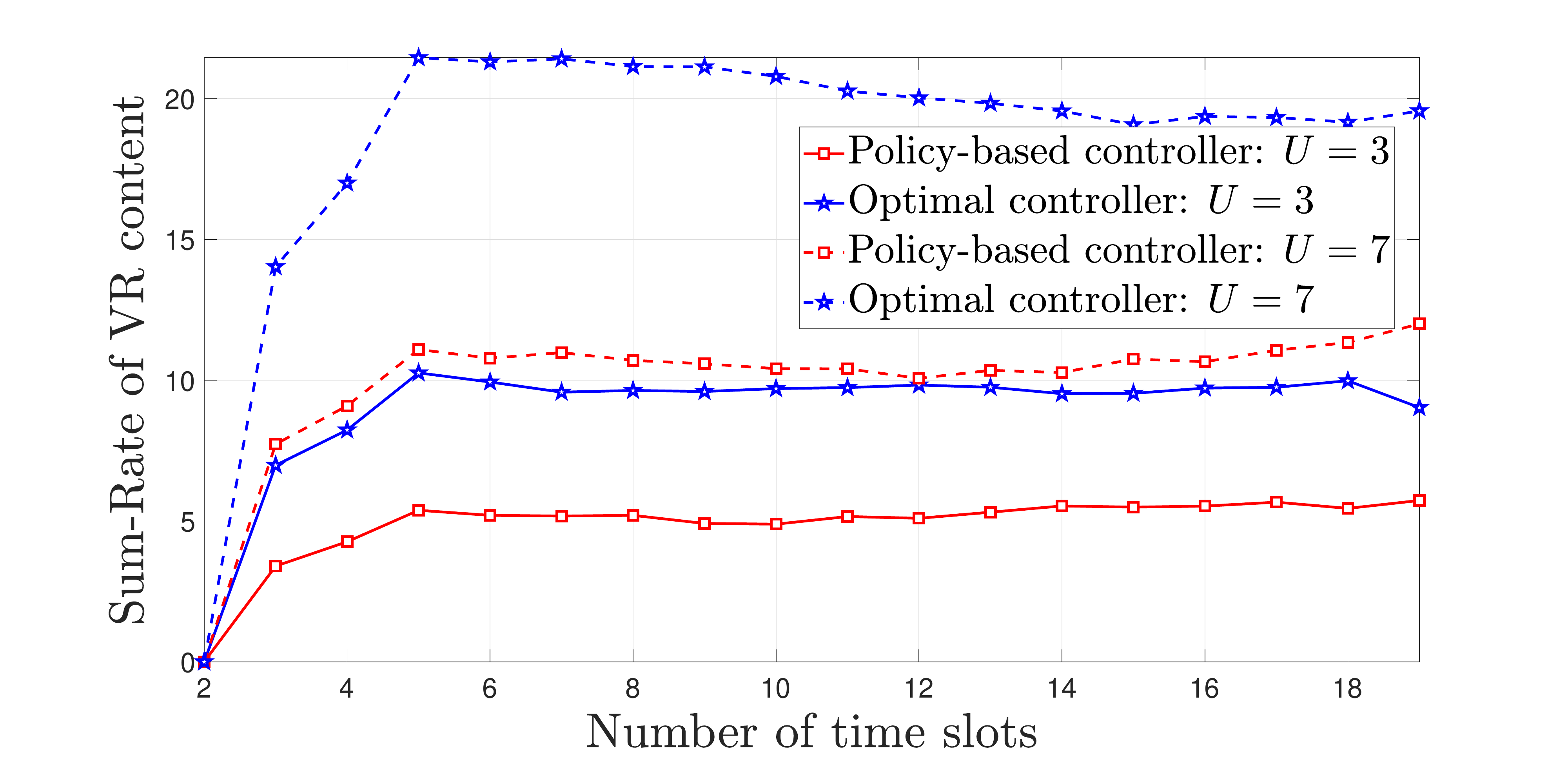}	
		\caption{ \small Sum-rate of \ac{VR} content vs. number of time slots.}
		\label{Sum_Rate}
	\end{center}
\vspace{-0.7cm}
\end{figure}
\indent We compare our \ac{RNN} \ac{RL} policy to the optimal solution for different number of users. In Fig.~\ref{Maximum_Q}, the maximum queue length over the number of iterations is plotted. Given that the \ac{RNN} was capable of capturing the dynamic behavior of the channel, for $U=7$, the maximum queue length for the policy-approach is only $0.52\%$ higher than the optimal maximum queue length. Meanwhile, for $U=3$ it is  nearly equal to the optimal solution.
Thus, this confirms that combining the risk-based approach with the \ac{RNN} \ac{RL} leads to highly reliable results. Clearly, the maximum queue length increases as the number of \ac{VR} users in the network increases.
Fig.~\ref{Sum_Rate} shows the sum-rate of \ac{VR} content, which constitutes the objective function in \eqref{opt2_problem}, over time. Here, the policy-based controller offers a solution that is considerably farther than the optimal solution in comparison to the results obtained for the maximum queue length in Fig.~\ref{Maximum_Q}. The reason for this is that a sum-term rate is being compared rather than individual rates; thus, the inaccuracy in every measure propagates into a higher inaccuracy when terms are added. As we can see, for $U=7$, the sum-rate for the policy-approach is $22\%$ less than optimal sum-rate. As for $U=3$, it is  $35\%$  less than the optimal solution. Clearly, as the number of \ac{VR} users increases, the sum-rate increases, and so does the gap between the optimal solution and the policy approach.
\vspace{-0.25cm}
\section{Conclusion}\label{Sec:Conclusion}
In this paper, we have investigated the problem of \ac{RIS}-\ac{VR} user association while guaranteeing reliable, low latency and high rate communications. We have proposed a risk-based aware optimization problem that takes into account the higher order statistics of the queue length, thus guaranteeing continuous reliability. The proposed problem was further transformed using Lyapunov optimization to a linear weighted function. Furthermore, the problem was solved using an \ac{RNN} \ac{RL} framework to reduce the dimensionality of the state space and capture channel dynamics and user mobility. 
\vspace{-0.15cm}
\bibliographystyle{IEEEtran}
\def\baselinestretch{0.86}
\bibliography{bibliography}
\end{document}